\documentclass[runningheads]{llncs}
\usepackage{todonotes}
\usepackage{comment}
\usepackage{makecell}
\usepackage{tabularx}
\usepackage{longtable}
\usepackage{booktabs,cellspace}
\usepackage{graphicx}
\usepackage{wrapfig}
\usepackage{amsmath}
\usepackage{tablefootnote}
\usepackage{hyperref}

\newcommand{\janik}[1]{\textcolor{black}{#1}}

\begin{document}
%
\title{Petri Net Classes for Collaboration Mining: Assessment and Design Guidelines}
\titlerunning{Petri Nets for Collaboration Processes}
%
\author{Janik-Vasily Benzin\inst{1} \and
Stefanie Rinderle-Ma\inst{1}}
\authorrunning{J.-V. Benzin and S. Rinderle-Ma}
%
\institute{Technical University of Munich, TUM School of Computation, Information and Technology, Garching, Germany\\
\email{\{janik.benzin,stefanie.rinderle-ma\}@tum.de} }
\maketitle              
\begin{abstract}
Collaboration mining develops discovery, conforman\-ce checking, and enhancement techniques for collaboration processes. 
The collaboration process model is key to represent the discovery result. As for process mining in general, Petri Net classes are candidates for collaboration process models due to their analytical power. However, a standard model class to represent collaboration processes is lacking due to the heterogeneity of collaboration and, thus, of collaboration mining techniques. Collaboration heterogeneity requires to cover, for example, intra-organizational collaborations as well as choreographies that span a process across multiple organizations. A standard collaboration model class would advance collaboration mining by focusing discovery through a common target model, supporting comparison, and enabling flexible mining pipelines. To find a standard model class, we aim at capturing collaboration heterogeneity in a meta model, assess Petri net classes as candidates for collaboration process models through the meta model, and derive design guidelines for the collaboration discovery. 
\end{abstract}

\section{Introduction}
\label{sec:intro}

Process mining research develops process discovery, conformance checking and enhancement techniques for \janik{\emph{process orchestrations}} that define what work is done in what order for similar \emph{cases} \cite{657557,jablonski1996workflow,aalst_service_2013}. In contrast, collaboration mining research develops the same for \emph{collaboration processes} that define what work is done in what order for \emph{collaborating cases}, i.e., collaboration processes correspond to multiple \janik{process orchestrations} that collaborate via departments \cite{liu_cross-department_2023}, services \cite{gaaloul_log-based_2008,stroinski_distributed_2019}, agents \cite{tour_agent_2021,tour_agent_2023,nesterov_discovering_2023}, and organizations \cite{zeng_cross-organizational_2013,zeng_top-down_2020,kwantes_distributed_2022,corradini_technique_2022}. 

For mining process orchestrations the de-facto standard model class is \emph{workflow nets} \cite{aalst_1999} to represent what work has to be done in what order as, for example, the workflow net concept underlies many models targeted by process discovery techniques \cite{augusto_automated_2019}. Although declarative process models were also proposed, they are by far in the minority \cite{augusto_automated_2019} and no discovery technique for collaboration processes targeting a declarative model is known to us. Hence, we focus on procedural models in the following. Nevertheless, a de-facto standard model class for collaboration processes is missing, as the model classes targeted by discovery techniques are diverse, e.g., \emph{communication nets} \cite{stroinski_distributed_2019} vs. composed \emph{RM\_WF\_nets} \cite{liu_cross-department_2023}. The heterogeneity of model classes is a consequence of the heterogeneity of  collaborations, e.g., \emph{message exchanges} \cite{corradini_technique_2022} and \emph{handover-of-work} \cite{tour_agent_2023}, from various perspectives, e.g., \emph{operational} \cite{stroinski_distributed_2019} vs. \emph{organizational} \cite{tour_agent_2023}, and on various granularity levels, e.g., intra-organizational \cite{liu_cross-department_2023} vs. inter-organizational \cite{corradini_technique_2022}.

The goal of this work is to identify potential standard model classes for collaboration processes as \textsl{``the choice of target model is very important for the discovery process itself''} \cite{van_der_aalst_representational_2011}. For this, we analyse the collaboration mining literature and related research areas with respect to the collaboration studied and integrate it in a \emph{collaboration process meta model} in \autoref{sec:system}. Next, we derive assessment criteria from the collaboration process meta model to assess standard Petri net classes with respect to required properties of a standard model class for collaboration processes and findings from the assessment in \autoref{sec:eval}. We focus on Petri net classes due to their expressiveness, graphical and integrating\footnote{Many alternative modelling languages can be transformed into Petri nets \cite{polyvyanyy_information_2019}.} nature, formal semantics, analysis techniques, and tool support \cite{peterson_petri_1977,aalst_1999,polyvyanyy_information_2019}. Similar to process mining, collaboration mining starts with discovering a collaboration process model \cite{aalst_service_2013} such that the targeted model class determines applicable conformance checking and enhancement techniques. Thus, we translate the assessment findings into design guidelines for process discovery in \autoref{sec:data} and conclude in \autoref{sec:conclusion}.

\section{A Collaboration Process Meta Model}
\label{sec:system}

The assessment of Petri net classes as standard candidates for collaboration process models in \autoref{sec:eval} necessitates an analysis of the heterogeneous collaboration studied in collaboration mining (cf. \autoref{sec:intro}), as the collaboration determines what must be modelled in a collaboration process model. Due to the heterogeneity in what is considered as a collaboration process \janik{(CP)}, we take an integrating approach to understand what has to be modelled in a collaboration process \janik{(CP)} model by presenting a collaboration process (CP) meta model. 

The meta model integrates specifications of what has to be modelled from research areas that study \janik{CP} models through a \emph{top-down} approach that starts from scratch and models the \janik{CP} as it should be executed. In that sense, collaboration mining represents the opposite \emph{bottom-up} approach by assuming the \janik{CP} has already been running for some time such that event logs can be extracted from information systems supporting the \janik{CP}. Hence, the CP meta model brings process models from both top-down and bottom-up approaches together and is presented in \autoref{ssec:elements} and instantiated for a real-world \janik{CP} in \autoref{sec:ex}.



\subsection{Elements of the Collaboration Process Meta Model}
\label{ssec:elements}

\begin{figure}
  \centering
  \includegraphics[width=\linewidth]{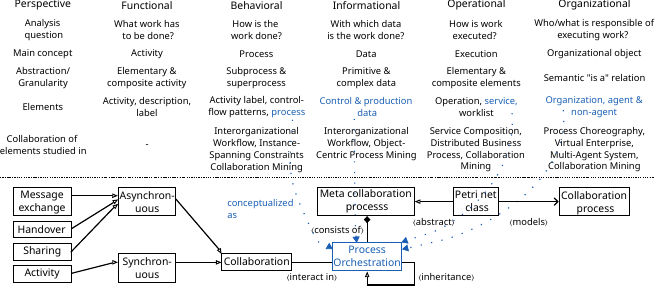}
  \caption{Process perspectives \cite{jablonski1996workflow} and research areas that study collaboration between perspective's elements and their relation to the collaboration process meta model.}
  \label{fig:all}
\end{figure}

To guide the integration of existing \janik{CP} models,
the functional, behavioral, informational, operational and organizational process perspectives \cite{jablonski1996workflow} are depicted in \autoref{fig:all}. Aside from the \janik{analysis} question defining the perspective on a process, the resulting main concept, the concept's abstraction relation instances/concept granularities, and elements of the perspective are presented. Existing work taking the top-down or bottom-up approach identifies a perspective's element as the answer to what is collaborating, i.e., the involved \emph{collaboration concepts} \janik{(perspective's elements denoted in blue in \autoref{fig:all})}.

Existing research areas taking the top-down approach are \emph{interorganizational workflows} \cite{657557,van_der_aalst_enacting_1999,van_der_aalst_p2p_2001}, \emph{process choreographies} \cite{alonso_local_2007,decker_interaction-centric_2011,fdhila_dealing_2015,meyer_automating_2015,fdhila_change_2015,fdhila_verifying_2022}, \emph{virtual enterprises} \cite{chu_partnership_2002,grefen_dynamic_2009,garcia_designing_2016}, \emph{distributed business processes} \cite{borkowski_event-based_2019,schulte_towards_2014}, \emph{service compositions} \cite{rinderle-ma_equivalence_2009,van_der_aalst_compositional_2009,aalst_service_2013}, \emph{multi-agent systems} \cite{tan_framework_2013,tour_agent_2021}, and (process) \emph{instance-spanning constraints} \cite{fdhila_classification_2016,winter_defining_2020}. In contrast, the bottom-up approach is taken by \emph{object-centric process mining} \cite{van_der_aalst_discovering_2020,barenholz_there_2023} and \emph{collaboration mining} \cite{gaaloul_log-based_2008,zeng_cross-organizational_2013,stroinski_distributed_2019,zeng_top-down_2020,corradini_technique_2022,kwantes_distributed_2022,nesterov_discovering_2023,liu_cross-department_2023,tour_agent_2023}. Considering both approaches, collaboration can be the result of multiple organizations (denoted in blue as an element of the organizational perspective in \autoref{fig:all}) or an organization's departments (non-agents in \autoref{fig:all}) collaborating towards achieving a common business goal. Also, collaboration can be the result of multiple services (element of the operational perspective) or agents (organizational perspective) collaborating to execute \janik{a CP}. \janik{Similarly, collaboration can be the result of multiple objects (informational perspective) collaborating to be processed in a CP}. Lastly, collaboration can be the result of multiple process instances (behavioral perspective) collaborating to meet the requirements stated in compliance constraints. 

Hence, the main element of the CP meta model defining \janik{the various collaboration concepts} can be an organization, (non-)agent, services, objects, and process instances. To cover the diverse nature of concepts, \janik{we conceptualize all these perspective's elements with their respective process orchestration, e.g., the process orchestration of a particular medical department \cite{liu_cross-department_2023} or of a particular web service \cite{stroinski_distributed_2019} (cf. ``conceptualized as'' relation in dotted blue in \autoref{fig:all}). Our process orchestration conceptualization in the CP meta model is in line with top-down and bottom-up approaches, as work for both approaches typically models its collaboration concept by some process orchestration, e.g., the underlying workflow net for each medical department \cite{liu_cross-department_2023} or for each web service \cite{stroinski_distributed_2019}. Hence, the \emph{meta collaboration process} that abstracts the Petri net classes modelling the CP in \autoref{fig:all} consists of process orchestrations that interact in various collaborations, i.e., the process orchestrations of a CP collaborate.}

Despite the diverse range of represented concepts in terms of process perspective and granularity level, the relationships between \janik{process orchestrations in the CP meta model (cf. \autoref{fig:all}) are either a collaboration, e.g., two agent's exchange messages, or \emph{inheritance} \cite{basten_inheritance_2001}, e.g., a service and its subservice are abstracted into a single service. The inheritance relation between process orchestrations captures the various granularity levels of existing works' collaboration concepts (department vs. organization) in the CP meta model.} The collaboration relationship can be generally defined as either asynchronous or synchronous \cite{657557} (cf. \autoref{fig:all}). \emph{Message exchanges}, e.g., http-messages in web services \cite{stroinski_distributed_2019}, \emph{handover-of-work} between \janik{process orchestrations}, e.g., agent A executes activity ``a'' and hands the work over to agent B that executes activity ``b'' \cite{tour_agent_2023}, and \emph{sharing}, e.g., a doctor in a clinic can only do one task at a time \cite{liu_cross-department_2023}, are studied for asynchronous collaboration between \janik{process orchestrations}. On the contrary, synchronous collaboration is studied as an \emph{activity} that is either executed by multiple \janik{process orchestrations} together, e.g., an internist and a surgeon consult to determine necessary medication for a patient \cite{liu_cross-department_2023}, or multiple \janik{process orchestrations} are referenced in an activity, e.g., \janik{the order and package for packing the items of an order \cite{van_der_aalst_discovering_2020}}. 

\janik{The applicability of the four collaboration types in the CP meta model to a real-world manufacturing \janik{CP} are shown in the next section.}

\subsection{Collaboration Process Meta Model Instance in Manufacturing}
\label{sec:ex}

\begin{figure}
  \centering
  \includegraphics[width=0.58\linewidth]{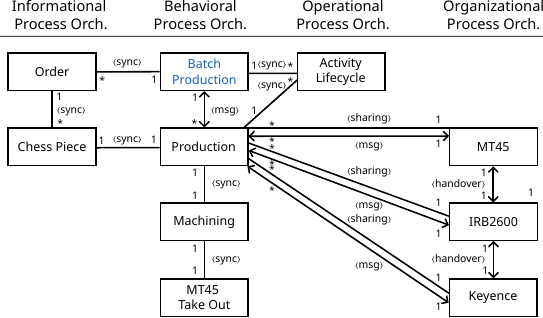}
  \caption{\janik{Selected} process orchestrations of a collaboration process meta model instance of the chess piece batch production recorded in \cite{mangler_xes_2023} as multiple event logs.}
  \label{fig:example}
\end{figure}

To show the applicability of the CP meta model, we instantiate it (depicted in \autoref{fig:example}) for a real-world manufacturing CP \cite{mangler_xes_2023}. The CP is a batch production of chess pieces orchestrated by the Cloud Process Execution Engine (CPEE) \cite{mangler_cloud_2022}. The CPEE executes the batch production \janik{process orchestration} (denoted in blue in \autoref{fig:example}) that refers to the top-level orchestration process executed by the CPEE. The batch production produces all chess pieces that are purchased in an order (informational \janik{process orchestration} related by synchronous collaboration \texttt{<sync>} in \autoref{fig:example}). The batch production process \janik{orchestration} instantiates a production \janik{process orchestration} for each produced chess piece by message exchange (\texttt{<msg>} relation between batch production and production) that is then executed in parallel by the CPEE. As soon as the production \janik{process orchestration} is finished, it sends a confirmation message back to the batch production \janik{process orchestration}. Cardinality constraints for the various collaboration relations are also depicted. For example, a single batch production produces all orders in the CP synchronously. 

Subprocesses instantiated by a behavioral \janik{process orchestration}, i.e., \janik{process orchestrations} instantiated with a blocking semantic \cite{mangler_cloud_2022}, are a synchronous collaboration, e.g., production and machining in \autoref{fig:example}. The CPEE has a fine-grained activity lifecycle model, that is referred to by the activity lifecycle operational \janik{process orchestration} in \autoref{fig:example}. Hence, the respective behavioral and the operational \janik{process orchestrations} synchronously collaborate by executing the respective activity lifecycle transitions. A MT45 lathe machine, an IRB2600 industrial robot and a Keyence precision measurement machine are responsible for executing the CP. As there is only a single MT45, IRB2600 and Keyence machine, these limited resources are shared by the production \janik{process orchestration}. The machines are orchestrated by message exchanges, i.e., their \janik{process orchestrations} collaborate in a \texttt{<msg>} collaboration with the production.

Following the meta model in \autoref{fig:all}, the CP meta model instance in \autoref{fig:example} abstracts a concrete Petri net CP model that can be mined for the batch production event logs. Depending on the \janik{analysis} question motivating the mining of a CP model for the batch production, a subset of \janik{process orchestrations} can be selected to scope the mined CP model. As it is not clear, what Petri net class should be targeted for the CP model by process discovery (cf. \autoref{sec:intro}), the next section assesses candidate Petri net classes.

\section{Assessment of Petri Net Classes in Modelling Collaboration Processes}
\label{sec:eval}

As elaborated in Sect. \ref{ssec:elements}, capturing CP in their entirety is a challenging task, in particular, w.r.t. the CP model class and the CP discovery technique which are tightly intertwined \cite{van_der_aalst_representational_2011}. Hence, in this section, we assess existing Petri Net classes for their capability of modelling and discovering CPs. 

\subsection{Assessment Criteria}
\label{ssec:criteria}

Given the existing research areas that study CPs in \autoref{sec:system}, through snowballing \janik{and expert knowledge}, we identify 20 Petri net classes that are target candidates as CP models for collaboration mining as depicted in \autoref{tab:models}. The assessment aims to determine the properties of each Petri net class in modelling CP models. The Petri net classes are presented in reference to their main study and year of publication. To assess the Petri net classes, \emph{general properties} characterize the class from a general conceptual and theoretical perspective, \emph{mining properties} characterize the class with respect to collaboration mining, and \emph{collaboration properties} characterize the class with respect to the four types of collaboration occurring in the CP meta model (cf. \autoref{fig:all}). 

\begin{table}[htb!]
    \centering
    \caption{Overview of the 20 Petri net model classes with CP modelling assessment. }
    \scalebox{0.77}{
    \begin{tabular}{lcccccccccccccc} 
    \toprule
    & & & \multicolumn{4}{c}{General Properties} & \multicolumn{4}{c}{Mining Properties} &  \multicolumn{4}{c}{Collaboration Properties} \\
    \midrule
     Class & \makecell{Main\\Study}  & Year & \makecell{Appr.}  &  \makecell{Turing\\Compl.} & \makecell{Data}  &  \makecell{Seman-\\tics} & \makecell{CP\\Repr.}  & \makecell{PD}   & \makecell{Redis-\\cover.} & \makecell{Sound-\\ness} & \makecell{Mes-\\sages} &  \makecell{Hand-\\over}   & \makecell{Shar-\\ing} & \makecell{Acti-\\vity} \\ 
    \midrule
    \makecell[l]{Petri Net}  & \cite{petri1962kommunikation} & 1962 &  T & $-$\cite{peterson_petri_1977}    & I & I/T  & I    & I & $-$ & NA  & I   & I  & I & I \\
    \makecell[l]{Colored Petri Net}  & \cite{jensen1981coloured} & 1981 &  T   & $-$\cite{jensen1981coloured}    & E & I  & I     & I & $-$ &  NA & E    & E  & E & E  \\
    \makecell[l]{Object System}  & \cite{valk_processes_1996} & 1996  &  T   & $+$\cite{kohler_properties_2004}   & E &  I & E      & I & $-$ & NA & E     & E   & E   & E   \\
    \makecell[l]{Interorgan.\\Workflow}  & \cite{657557} & 1998 & T    & $-$\cite{peterson_petri_1977}  & I   & I & E    & I& $-$ &   D & E   & I   & I    & E    \\
    \makecell[l]{Interaction\\Petri Net}  & \cite{alonso_local_2007} & 2007 & T   & $-$\cite{peterson_petri_1977} & I & I & E   & \makecell[c]{\cite{gaaloul_log-based_2008}\\\cite{corradini_technique_2022}}  & $-$ &  ?  & E\tablefootnote{Interaction Petri nets are the only model class in our overview, that conceptualizes a message exchange as an atomic, synchronous firing of a transition in a Petri net \cite{alonso_local_2007}.}  & I  & I    & $-$ \\
    \makecell[l]{Compositional\\Service Tree}  & \cite{van_der_aalst_compositional_2009} & 2009 & T  & $-$\cite{peterson_petri_1977} & I  & I &  E     & ? & $-$ & D & E   & I   &  I  & $-$ \\
    \makecell[l]{$\upsilon$-PN}  & \cite{rosa-velardo_decidability_2011} & 2011 & T   & $-$\cite{rosa-velardo_decidability_2011} & E  & I & I   &  ? & $-$ &   U & I  & I  & I  & I  \\
    \makecell[l]{Integrated\\RM\_WF\_net}  & \cite{zeng_cross-organizational_2013} & 2013 &  B & $-$\cite{peterson_petri_1977} & I  & I  & E      &  \cite{zeng_cross-organizational_2013} & $-$ &  D & E     & I & E   & E  \\
    \makecell[l]{Healthcare\\Petri Net}  & \cite{mahulea_modular_2018} & 2018 &  T & $-^?$ & I  & I  & E      &  ? & $-$ &  ? & I     & E & E   & $-$  \\
    \makecell[l]{Synchronuous\\Proclet System}  & \cite{fahland_describing_2019} & 2019 & T   & $-^?$\cite{fahland_describing_2019} &  E  & T & E    &  ? & $-$ &   NA & I    & I   & I  & E   \\
    \makecell[l]{t-PNID}  & \cite{polyvyanyy_information_2019} & 2019  & T  & $-$\cite{peterson_petri_1977} & E  & I & E     & I & $-$ &  U  & I   & I  & I & E   \\
    \makecell[l]{Communication\\Net}  & \cite{stroinski_distributed_2019} & 2019 & B  & $-^?$ & I  & I & E      &  \cite{stroinski_distributed_2019} & $-$ &  ? & E  & I  & I  & $-$   \\

    \makecell[l]{Top-Level\\Process Model}  & \cite{zeng_top-down_2020} & 2020 & B  & $-^?$ & I  & I & E    & \cite{zeng_top-down_2020} & $-$ &   D  & E   &   I  &   I &  I  \\
    \makecell[l]{System Net}  & \cite{reisig_composition_2020} & 2020 & T & $-^?$ & E  & T  & E    & ? & $-$ &   NA & E  & E  & I  &  I  \\
    \makecell[l]{Object-centric\\Petri Net}  & \cite{van_der_aalst_discovering_2020} & 2020 & B  & $-^?$ & E  & I  & E   & \cite{van_der_aalst_discovering_2020} & $-$ &   ?   & I    & I  & I  & E \\
    \makecell[l]{Industry Net}  & \cite{kwantes2018synthesis} & 2022 & T  & $-^?$ & I  & I &  E   &  \cite{kwantes_distributed_2022} & $-$ &   ? & E  & I  &  I   & $-$ \\
    \makecell[l]{Multi-Agent\\System Net}  & \cite{tour_agent_2023} & 2023 & B  & $-$\cite{peterson_petri_1977} & I   & I & E     &  \cite{tour_agent_2023} & $-$ &  D  & I   & E &  I  & I  \\
    \makecell[l]{Generalized\\Workflow Net}  & \cite{nesterov_discovering_2023} & 2023 & B & $-^?$ & I   & I  &  E  & \cite{nesterov_discovering_2023} & $-$ &  D   & E     & I  & I   & E  \\
    \makecell[l]{Composed\\RM\_WF\_net}  & \cite{liu_cross-department_2023} & 2023 & B & $-$\cite{peterson_petri_1977} & I   & I  &  E  & \cite{liu_cross-department_2023} & $-$ &  ?   & E     & I  & E   & E  \\
    \makecell[l]{Typed Jackson\\Net}  & \cite{barenholz_there_2023} & 2023 & B   & $-$\cite{barenholz_there_2023} & E & I & E  &  \cite{barenholz_there_2023} & $+$  &   S  & I   & I  &  I  & E  \\

    \bottomrule 
    \end{tabular}}
    \label{tab:models}
\end{table}

General properties of Petri net classes are the \emph{approach} \cite{meyer_automating_2015}, \emph{Turing completeness} \cite{peterson_petri_1977}, ability to represent \emph{data} (cf. control vs. production data in \autoref{sec:system}) and the proposed type of \emph{semantics} \cite{van_glabbeek_petri_1987}. To build a CP model, two approaches exist: \emph{(T)op-down} starts from scratch and models the CP manually by starting with the definition of \janik{process orchestrations} and collaborations and further refining the workflow nets, while \emph{(B)ottom-up} starts with an event log recorded from information systems supporting the CP execution and mines a CP model. Next, Petri net classes can either be Turing complete ($+$) or not ($-$). If the Turing completeness is only conjected, the conjecture is denoted with a ?, e.g., $-^?$. 

As the ability to represent a concept in a Petri net class can either be $-$, i.e., the class definition prohibits the representation, \emph{(I)mplicit}, i.e., the concept can theoretically be represented in the class, but has no dedicated element in the definition, or \emph{(E)xplicit}, i.e., the concept has a dedicated element in the definition, the properties data, \emph{CP representation} and all four collaboration properties are based on this distinction. Petri net semantics can either be \emph{(I)nterleaving}, i.e., semantics is defined as traces of fired transitions such that transition labels always interleave/are totally ordered, and \emph{(T)rue concurrency}, i.e., semantics is defined as a \emph{causal net} \cite{van_glabbeek_petri_1987} such that transition firings can be partially ordered or "truly concurrent". 

Mining properties are the CP representation, \janik{\emph{discovery technique (PD)}}, \emph{rediscoverability} \cite{van_der_aalst_workflow_2004}, and \emph{soundness} \cite{657557}. As CPs are a composition of \janik{process orchestrations}, e.g. \cite{657557,liu_cross-department_2023}, CP representation refers to the representation of workflow net compositions. To mine a CP model that is an instance of the Petri net class, a discovery technique targeting that Petri net class has to either exist, is unknown to exist (\emph{?}), or is \emph{(I)nherited} from a Petri net subclass. An important and desired property of a Petri net \janik{for collaboration mining} is its soundness (and generalized definitions such as \emph{identifier soundness} \cite{barenholz_there_2023}) that can either be \emph{(N)ot (A)pplicable} as the class is too general, unknown (\emph{?}) as a restated definition is missing, \emph{(U)ndecidable} or \emph{(D)ecidable} as a decision problem, and \emph{(S)ound by construction}. 

\subsection{Findings}
\label{ssec:findings}

From the assessment result in \autoref{tab:models}, we deduce nine findings. The first finding is that research on candidate Petri net classes started with the top-down approach and has gradually put more focus on the bottom-up approach. The second finding is that for 19 Petri net classes, the approach assessment is straightforward, while the typed Jackson net proposal \cite{barenholz_there_2023} fuses the top-down with the bottom-up approach by taking both approaches in turn and bringing them together through a framework for rediscoverability. However, their motivation puts more focus on the bottom-up approach. The third finding is the observation that Typed Jackson nets and object-centric Petri nets \cite{van_der_aalst_discovering_2020} are the only bottom-up approaches with an explicit data representation, a consequence of recent efforts to mine CP models for the informational process perspective (cf. \autoref{fig:all}) in which \janik{process orchestrations} are objects and collaboration is synchronous. Considering the general properties, a fourth finding is that there is no existing \janik{discovery technique} that defines a true concurrency semantics for its Petri net class. The fourth finding contrasts a recent effort on true concurrency through \emph{systems mining} by \cite{fettke_systems_2022} and related efforts on partial order-based process mining \cite{leemans_partial-order-based_2022}. 

Considering the mining properties assessment in \autoref{tab:models}, a fifth finding is that early Petri net classes inherit discovery techniques from later proposals for Petri net subclasses, e.g. the \emph{Colliery technique} \cite{corradini_technique_2022} is a discovery technique for the Interaction Petri net class, which is a subclass of the interorganizational workflow proposed nine years earlier. The sixth finding is that discovery techniques were only proposed \janik{three times} for an existing Petri net class proposed in a top-down approach, i.e., the two approaches are brought together through distinct publications in \cite{gaaloul_log-based_2008,corradini_technique_2022} for the interaction Petri net and in \cite{kwantes_distributed_2022} for the industry net. We conject that both the interaction Petri net and industry net are equivalent both in terms of the assessement and theoretically, which emphasizes the need for a central Petri net class for collaboration mining to alleviate potential redundancies in research. The seventh finding is that rediscoverability and sound by construction are rare and coincide in typed Jackson nets, while the importance of soundness in mining is confirmed through classes having a \janik{discovery technique}, soundness is either \emph{?}, decidable or sound by construction. 

Considering the collaboration properties assessment in \autoref{tab:models}, the eighth finding is that colored Petri nets and object systems are the only classes that explicitly represent all four collaboration types, but lack a native discovery technique targeting that class. Although the discovery technique \cite{van_der_aalst_discovering_2020} for object-centric Petri nets results in an inherited discovery technique for colored Petri nets, the former class lacks explicit representation of three of the four collaboration types. In contrast, the classes integrated RM\_WF\_net and composed RM\_WF\_net targeted by the discovery techniques in \cite{zeng_cross-organizational_2013} and \cite{liu_cross-department_2023} explicitly represent three of the four collaboration types. The multi-agent system net targeted by the agent miner \cite{tour_agent_2023} is the only class particularly designed for explicitly representing the missing handover collaboration type, pointing to a potential fused class with explicit representation of all four classes and a discovery technique that comes with it. Overall, the eighth finding highlights that the early classes colored Petri net and object systems are too general and powerful to be targeted by a discovery technique (yet), but proposed subclasses for which a discovery technique exists do not yet come with the explicit modelling of all four collaboration types. The ninth and final finding is that there is yet no class for collaboration mining that the workflow net constitutes for process orchestration mining. The next section translates the nine findings into design guidelines for collaboration miners. 
 
\section{Design Guidelines for Process Discovery}
\label{sec:data}

The assessment of existing Petri net classes as standard candidates for CPs \autoref{sec:eval} resulted in nine findings that are the foundation of the design guidelines for future process discovery in collaboration mining. The search for the standard Petri net class for collaboration mining continues and underlies the first three guidelines: 

\begin{enumerate}
    \item[\textsl{G1}] Take the balance between (explicit) modelling and decision power of your targeted Petri net class actively into account. Existing \janik{discovery techniques} target either the one end, e.g. \cite{barenholz_there_2023} with good decision power on soundness and rediscoverability, but lack of three collaboration types, or the other end of the bargain, \cite{liu_cross-department_2023} with good modelling power, but lack of known decision power on soundness and rediscoverability. A future discovery technique targeting a (potentially new) class that has modelling power to represent all four collaboration types and good decision power not only brings together the top-down and bottom-up approach, but also is applicable to the challenges studied in a diverse range of research areas and their respective application domains (cf. \autoref{fig:all}). 
    \item[\textsl{G2}] Match your discovery technique's design rationale with the properties of existing classes to access the existing theoretical and practical knowledge and improve rather algorithmically instead of by definition of a new class. If a new class is still necessary, take the first guideline into account.
    \item[\textsl{G3}] State your targeted class explicitly and formally, as it allows straightforward assessment similar to \autoref{ssec:criteria} and is part of the first maturity stage \cite{van_der_werf_all_2023} of your discovery technique. Have all four maturity stages in mind. 
    \item[\textsl{G4}] Think three times about not taking a divide-and-conquer approach to discovery, as the CP is a composition. If you have another approach, motivate it well. 
    \item[\textsl{G5}] Balance your design guided by the previous four guidelines with your discovery technique's complexity to achieve sufficient efficiency for practical purposes. 
\end{enumerate}

\janik{By taking these guidelines into account, collaboration mining techniques can be flexibly combined, easily compared and assessed, and come with a well-researched theoretical foundation.}

\noindent\textbf{Limitations:} Despite the extensive set of covered techniques and research areas, the heterogeneity of collaboration results may result in having missed existing work on collaboration. Also, the set of assessed Petri net classes is likely not completely comprehensive. Hence, a Petri net class might exists already that is the best candidate for a standard CP model. \janik{Additionally, the criteria can be further refined to improve the characterization of the standard CP model.}

\section{Conclusion}
\label{sec:conclusion}

The integration of heterogeneous collaboration in our CP meta model enables the assessment of existing Petri net classes with respect to properties of a standard model class for CP and can guide new collaboration mining techniques beginning with process discovery. As a clear standard Petri net class for CP is still missing, future techniques should take this lack of standardization into account to advance collaboration mining in a coherent way. Additionally, a fusion of the top-down and bottom-up approach opens up yet to be utilized ways of combining results from both sides.

\bibliographystyle{splncs04}
\bibliography{main}
\end{document}